\documentclass{epl}

\title{Noncommutative fields in three dimensions and mass generation}
\shorttitle{Noncommutative fields}
\author{J. R. Nascimento \inst{1} \and A. Yu. Petrov \inst{1} \and R. F. Ribeiro\inst{1}}

\institute{                    
  \inst{1} Departamento de Fisica, Universidade Federal da Paraiba, Caixa Postal 5008, 58051-970, Jo\~{a}o Pessoa, Paraiba, Brasil
}
\pacs{11.10.Nx}{Noncommutative field theory}

\begin{document}

\maketitle

\begin{abstract}
We apply the noncommutative fields method for gauge theory in three dimensions where the Chern-Simons term is generated in the three-dimensional electrodynamics. Under the same procedure, the Chern-Simons term turns out to be cancelled in the Maxwell-Chern-Simons theory for the appropriate value of the noncommutativity parameter.
Unlike of the four-dimensional space, in this case the Lorentz symmetry breaking is not generated. 
\end{abstract}


\newpage
\section{Introduction}

The concept of noncommutativity has deep motivations originated
from the fundamental properties of the space-time and string
theory \cite{SW}. One of the important implications of the
noncommutativity is the Lorentz violation \cite{Car}
which, in part, modifies the dispersion
relations \cite{Gamboa1,Mag}. The new mechanism of implementing noncommutativity which could provide
breaking of the Lorentz invariance, that is, the noncommutative
fields method, was suggested in
\cite{Gamboa1}. Its essence consists in the fact that the
noncommutativity affects not the space-time coordinates as it takes place
in the typical noncommutative field theory models but the canonical variables
of the theory, that is, their algebra receives a noncommutative deformation which was shown to have some important applications, for example, natural explication of matter-antimatter asymmetry.
As a result, the Hamiltonian of the theory (and, as a consequence,
the Lagrangian after reformulation of the theory in terms of the
Lagrange variables) turns out to be modified by new additive Lorentz-breaking
terms. In the papers \cite{Gamboa2,Gamboa3} this method was
applied to the four-dimensional electrodynamics and to the Yang-Mills theory
where new gauge invariant terms in the Lagrangian were generated. 

In this works we are interested in applying  the method showed in \cite{Gamboa1} to induction of the
Chern-Simons action in (2+1)-dimensional spacetime. So by consequence we generate topological mass 
term in the gauge theory without breaking gauge symmetry and without coupling to the fermionic matter which is the most common way to introduce mass to gauge field with preservation of gauge invariance.
There is a well known class of (2+1)-dimensional spacetime theories exhibiting interesting
phenomena such as exotic statistics, fractional spin and massive gauge fields \cite{DJT}.
All these phenomena are of topological nature and they can be produced when we add Chern-Simons
term to the Lagrangian which describe the system under consideration.
In general the Chern-Simons term is generated by radiative correction when the gauge field 
is coupled to the fermionic field.  This effect has been studied by Redlich \cite{Red} in the context of the quantum electrodynamics in three-dimensional space-time. Proceeding along the lines \cite{Red} other quantum field theory models were investigated \cite{mva, mg, ro}.  Extension to higher odd dimensions was carried out including the case of a gravitational background field \cite{ADM, V}.

In Section II, we apply the noncommutative fields method to generate the Chern-Simons term to the pure Maxwell theory. A comment and a conclusion on our results comprises the Section III. 

\section{Pure Maxwell Theory}
The starting point now is the simple Maxwell action which after
splitting the indices into time (zero) and space
(denoted by Latin letters) ones, and taking the Minkovski metric $\eta_{\mu\nu}=diag(-++)$ gets the form
\begin{eqnarray}
\label{lagr10}
S=\int d^3x\Big(-\frac{1}{4}F_{ij}F_{ij}+\frac{1}{2}(\dot{A}_i-\partial_i A_0) (\dot{A}_i-\partial_i A_0)\Big).
\end{eqnarray}

First we find the momenta: 
\begin{eqnarray} 
p_i=\frac{\partial
L}{\partial\dot{A}_i}=\dot{A}_i-\partial_i A_0\equiv F_{0i}, 
\end{eqnarray} 
from which we express
\begin{eqnarray} 
\dot{A}_i=p_i+\partial_i A_0. 
\end{eqnarray} 
The action
(\ref{lagr10}) does not depend on $\dot{A}_0$, hence we have the
primary constraint $\Phi^{(1)}=p_0=0$. The Hamiltonian is as usual
\begin{eqnarray} 
H=p_i\dot{A}_i-L=\frac{1}{2}p_ip_i-A_0\partial_i
p_i+\frac{1}{4}F_{ij}F_{ij}, 
\end{eqnarray} 
with the secondary constraint is
$\Phi^{(2)}\equiv \Delta=-\partial_i p_i$. This constraint, from the Hamiltonian viewpoint, after transformation of
canonical variables to operators generates the gauge
transformations by the rule 
\begin{eqnarray} 
\label{gat}
\delta A_i(\vec{x})&=&\frac{1}{i}[A_i(\vec{x}),\int d^2
\vec{x}'\xi(\vec{x}')\Delta(\vec{x}')]=\partial_i\xi(\vec{x});\nonumber\\
\delta p_i(\vec{x})&=&\frac{1}{i}[p_i(\vec{x}),\int d^2
\vec{x}'\xi(\vec{x}')\Delta(\vec{x}')]=0, \end{eqnarray}
which reproduces the gauge transformations following from the
Lagrangian formulation of the theory. 

Now let us implement noncommutative field method which consists in modifying the 
fundamental commutation relations to the following form: 
\begin{eqnarray}
&&[A_i(\vec{x}),A_j(\vec{y})]=0;\nonumber\\
&&[p_i(\vec{x}),p_j(\vec{y})]=i\theta_{ij}\delta(\vec{x}-\vec{y});\nonumber\\
&&[A_i(x),p_j(y)]=i\delta_{ij}\delta(\vec{x}-\vec{y}). 
\end{eqnarray} 
The form of the gauge transformations (\ref{gat}) will be unchanged if the
generator $\Delta$ is deformed as 
\begin{eqnarray} 
\Delta=-\partial_i
p_i+\theta_{ij}\partial_iA_j. 
\end{eqnarray} 
Consequently, the Hamiltonian will
be modified. Its form turns out to be 
\begin{eqnarray}
H'=\frac{1}{2}p_ip_i+\frac{1}{4}F_{ij}F_{ij}+A_0(-\partial_i
p_i+\theta_{ij}\partial_iA_j). 
\end{eqnarray} 
The equations of motion are 
\begin{eqnarray}
\dot{A}_i&=&p_i+\partial_iA_0;\nonumber\\ \dot{p}_i&=&-\partial_j
F_{ji}+\theta_{ij}\partial_i A_0. 
\end{eqnarray} 
The canonical variables in the
theory are $\pi_i$ satisfying the commutation relations 
\begin{eqnarray}
&&[\pi_i(\vec{x}),\pi_j(\vec{y})]=0;\nonumber\\
&&[A_i(x),\pi_j(y)]=i\delta_{ij}\delta(\vec{x}-\vec{y}). 
\end{eqnarray}
Their form is 
\begin{eqnarray} 
\pi_i=p_i-\frac{1}{2}\theta_{ij}A_j. 
\end{eqnarray} 
The corresponding deformed Lagrangian looks like 
\begin{eqnarray} 
L'=\pi_i \dot{A}_i-H',
\end{eqnarray} 
which manifest form is 
\begin{eqnarray}
L'=-\frac{1}{4}F_{\mu\nu}F^{\mu\nu}-\frac{1}{2}\theta_{ij}(\dot{A}_i
A_j+A_0 F_{ij}). 
\end{eqnarray} 
It is very natural to suggest
$\theta_{ij}\equiv\theta\epsilon_{ij}$, where $\theta$ is a scalar
noncommutativity parameter. In this case, the Lagrangian takes the
form 
\begin{eqnarray}
\label{deflan}
L'=-\frac{1}{4}F_{\mu\nu}F^{\mu\nu}+\frac{\theta}{2}\epsilon^{\mu\nu\lambda}A_{\mu}\partial_{\nu}A_{\lambda}.
\end{eqnarray} 
We find therefore that for the three-dimensional pure Maxwell
theory the noncommutative deformation of the canonical variables
algebra leads to generation of the Chern-Simons term!

At the same time, the straightforward repeating of this procedure for the Maxwell-Chern-Simons theory \cite{DJT}
\begin{eqnarray}
\label{lagr}
S&=&\int d^3
x\Big(-\frac{1}{4}F_{\mu\nu}F^{\mu\nu}+m\epsilon^{\mu\nu\lambda}A_{\mu}\partial_{\nu}A_{\lambda}
\Big)
\end{eqnarray}
implies in the following analog of the deformed Lagrangian (\ref{deflan}):
\begin{eqnarray} 
\label{act2}
L'=-\frac{1}{4}F_{\mu\nu}F^{\mu\nu}+(m+\frac{\theta}{2})\epsilon^{\mu\nu\lambda}A_{\mu}\partial_{\nu}A_{\lambda}.
\end{eqnarray} 
Thus, we conclude that within the
framework of the noncommutative field method the
Maxwell-Chern-Simons theory is converted to the Maxwell-Chern-Simons theory with shifted mass. In particular, for $\theta=-2m$, it is converted to the common electrodynamics.

\section{Comment and Conclusion}

The results obtained allow us to make three important conclusions.
First, unlike of the four-dimensional case \cite{Gamboa2,Gamboa3}, in three space-time dimensions the noncommutative deformation of the canonical operator algebra does not generate breaking of the Lorentz symmetry. It is quite natural since the noncommutative field method applied to gauge theories in \cite{Gamboa2,Gamboa3} implies in arising of Chern-Simons-like gauge invariant term which in four-dimensional case, unlike of three dimensions, cannot be constructed without introducing of a Lorentz-breaking constant vector. Second, under applying of new commutation relations for momenta the Maxwell theory is converted to the Maxwell-Chern-Simons theory. Third, the remarkable result consists in arising of a new mechanism of topological mass generation (see the ref. \cite{JaDva} for discussion) which however does not require coupling to matter. 

We also found that this procedure can be straightforwardly applied for the non-Abelian case. Developing this method for the three-dimensional Yang-Mills theory, one can see that the pure Yang-Mills action
\begin{eqnarray}
S_{YM}=\frac{1}{2g^2}{\rm tr}F_{\mu\nu}F^{\mu\nu}
\end{eqnarray}
receives the correction
\begin{eqnarray}
\Delta S=\theta\epsilon_{\mu\nu\lambda}(A^{\mu}F^{\nu\lambda}+\frac{2}{3}A^{\mu}A^{\nu}A^{\lambda}),
\end{eqnarray}
which reproduces the well-known structure of the noncommutative Chern-Simons term (note that the similar term in four dimensions, which, however, violated the Lorentz symmetry, was obtained in \cite{Gamboa3}). The details of calculation will be given elsewhere.

\acknowledgements This work was partially support by CNPQ, CNPQ/PROCAD, CNPQ/FINEP/PADCT and PRONEX/CNPQ/FAPESQ. A. Yu. P. has been supported by the CNPq/FAPESQ DCR program (CNPq project 350400/2005-9).

\end{document}